\documentclass[aps,prx,twocolumn,a4paper]{revtex4-2}

\usepackage{hyperref}
\usepackage{geometry}                		
\usepackage{graphicx}				
\usepackage{amsmath}
\usepackage{amssymb}
\usepackage{color}

\usepackage{ascii}
\usepackage{tikz}
\usetikzlibrary{quantikz2}
\usepackage{adjustbox}

\usepackage{algorithm}
\usepackage{algpseudocode}

\begin{document}

\title{Neural Guided Sampling for Quantum Circuit Optimization}
\author{Bodo Rosenhahn$^1$, Tobias J.\ Osborne$^2$, Christoph  Hirche$^1$}
\affiliation{1) Institute for Information Processing (tnt/L3S), Leibniz Universit\"at Hannover, Germany\\
\{rosenhahn, hirche\}@tnt.uni-hannover.de}
\author{ }
\affiliation{2) Institute for Theoretical Physics (ITP/L3S), Leibniz Universit\"at Hannover, Germany\\
tobias.osborne@itp.uni-hannover.de}

\date{\today}

\begin{abstract}
Translating a general quantum circuit 
on a specific hardware topology with a reduced set of available gates, also known as transpilation,
comes with a  substantial increase in the length of the equivalent circuit. Due to decoherence, the quality of the computational outcome can degrade seriously with increasing circuit length. Thus, there is major interest
to reduce a transpiled quantum circuit to an equivalent circuit 
which is in its gate count as short as possible. This is, what we call quantum circuit reduction: Finding a quantum circuit with a reduced amount of gates without changing its unitary.
One method to address efficient transpilation, e.g. as a post-transpilation process, is
based on approaches known from stochastic optimization, e.g. by using random sampling and local resynthesis strategies. 
Here, a core challenge is that these methods can suffer from sampling efficiency, causing long and energy consuming optimization time. As a remedy, we propose in this work  2D neural guided sampling. Thus, given a 2D representation of a quantum circuit, a neural network predicts groups of gates in the quantum circuit, which are likely reducible. It leads to a  sampling prior which can heavily reduce the compute time for quantum circuit reduction.
In several experiments, we demonstrate that our method is superior to results obtained from different qiskit or BQSKit optimization levels.  
\end{abstract}

\maketitle

\section{Introduction}
The task of translating a given quantum circuit into a sequence of gates realizable on a provided hardware architecture is generally known as \emph{Quantum Architecture Mapping} (QAM) \cite{Datta22, Stefano2024} or quantum circuit transpilation.
 Due to decoherence and the overhead of hence necessary quantum error correction, it is crucial to optimize for efficient and short quantum circuits to ensure reliable results when they are realized in hardware. 
 The quantum volume metric \cite{Ryabov15} is a prominent score for evaluating the capabilities and error rates of a quantum computer and is an important measure to compare the realization of quantum circuits on different hardware. 
 Efficient transpilation algorithms have a direct impact on the quantum volume metrics and are of huge interest for quantum chip developers. 
 It is of a similar importance to electronic design automation (EDA) \cite{Top23} for the digital domain. 
Naive transpilation replaces each general quantum gate with a set of hardware realizable quantum gates. Afterwards optimizers and rule-sets are used to reduce the length of the quantum circuit. E.g. if two gates realize an identity mapping, they can be removed. Such approaches for cleaning the quantum circuit can be seen as post-transpilation. They are typically part of the transpilation itself (e.g. driven by different optimization levels). Therefore, we assume a naively (inefficiently) transpiled gateset, or a quantum circuit which is constructed from a set of realizable base gates. In this work, we propose optimizing approaches to achieve a more efficient representation of this circuit, measured in a reduced circuit length and depth.    
The process of automatically finding the best quantum circuit to implement a specific target unitary is sometimes referred to as \emph{quantum architecture search} (QAS).
This name is inspired by terminology from the machine learning community, where neural architecture search \cite{Miikkulainen2020,xie2018snas} involves selecting algorithms and tuning their hyperparameters.
 Common methods include reinforcement learning, structural search, genetic algorithms or performance prediction, as described in \cite{BakerGRN18,CaiCZYW18,negrinho2019towards,ARUFE2022101030}.
To address the challenge of minimizing the length of transpiled  quantum circuits, we use in this work a term \textit{replacement scheme} which performs local resynthesis. It acts (in contrast to earlier work) on the 2D graphical quantum circuit model. 
This work proposes  a method based on 
a neural network to predict a likelihood map on the 2D quantum circuit representation to identify blocks that are likely reducible. 
The likelihood map can then be used for guided sampling and should increase the sampling efficiency drastically, as unlikely areas for circuit reduction are avoided. 
This method, known as neural guided sampling, has been successfully applied to other fields such as multi-model fitting \cite{9008398,KluBra2020,9010674}. We utilize this approach for optimizing quantum circuits.

 Our core contributions can be summarized as follows
\begin{enumerate}
\item  We propose a 2D random search (RS) based term replacement algorithm to identify reducible quantum circuit sub-blocks in a quantum circuit. These local replacements perform local resynthesis. Iterative block replacements reduce the overall quantum circuit length, until a stopping criteria has been met. 
\item We extend sampling from an equal distribution by using a neural network to identify sub-blocks in the 2D quantum circuit representation which are likely reducible. It is a regression network that predicts a likelihood map which is used for neural guided sampling (NGS).
\item In several experiments, we demonstrate that our approach is faster compared to other iterative optimization schemes and leads to more efficient quantum circuits. It is also superior  in terms of achieved circuit length compared to different qiskit and BQSKit optimization levels.
\end{enumerate} 
Please note, that our optimization approach is reasonably slow, compared to existing rule based transpilation schemes. Instead, our aim is to find optimized circuits with minimal gate count and depth. With regards to the practical utility, this can be interesting, when a circuit is used very often, e.g. in a hybrid optimization setting. Our approach can also be interesting for frequently used building blocks,  e.g. a phase estimation component, a diffusor block or the realization of a convolution.  

The remainder of this paper is structured as follows: 
Section II starts with the literature overview. Section III presents fundamentals on random search  and summarizes computer graphs and our main competitor, the 1D random sampling based quantum circuit optimization approach \cite{rosenhahn2025optimization}. Afterwards we introduce the 2D  neural guided sampling based quantum circuit optimization. In Section IV we perform several experiments to demonstrate that our methods achieve (a) more efficient quantum circuits (code with less gates to represent
the same unitary) and (b) the optimization of 
the reduced circuit is faster.

We decided to perform experiments on two gate sets. The first gate set we use is common for ion-trap architectures and comprises $RX,RY,RZ$  and $RXX$ gates. The
second gate set we use is common for NISQ-architectures and comprises $RX, RZ$ and $CZ$ gates.
Section V summarizes our paper and gives a brief outlook to further work.
 
\section{Literature overview}
Recent work addresses the challenge that many quantum transpilers are proprietary and  optimized for dedicated hardware. In \cite{kundu2025}   compiler confidentiality is approached using reverse-engineering compilation methodologies. There a simple ML-based framework is used to infer underlying optimization techniques by leveraging structural differences observed between original and compiled circuits. A neural network detects optimization passes which demonstrate the viability of this threat to compiler confidentiality. In \cite{ren2024tackling}   a cross-layer approach for coherent error mitigation is proposed. The authors consider program-level, gate-level, and pulse-level compiler optimizations, by leveraging the hidden inverse theory and by exploiting structured information. The work \cite{wang2025tetris}
 addresses the challenge that untrusted quantum compilers pose significant risks as they can lead to the theft of quantum circuit designs and compromise sensitive intellectual property. Therefore, they 
 propose TetrisLock, a split compilation method for quantum circuit obfuscation that uses an interlocking splitting pattern. In \cite{Younis22}, the authors 
 describe algorithms to apply parameterized circuit instantiation during  circuit optimization and gate-set transpilation. The authors propose an iterative scanning gate removal with the circuit being  iteratively scanned by using a linear strategy.
The work \cite{huo2025revisiting} revisits the necessity of frequent noise-adaptive transpilation and concludes  that the classical overhead associated with per-circuit noise-aware transpilation may not always be justified. Large language models to enhance the transpilation process have been proposed in \cite{siavash2025LLM}, but the study reveals that LLMs are performing badly at this task.
 The transpilation of quantum circuits for execution of quantum algorithms across different cores while minimizing inter-core communications has been investigated in \cite{10.1145/3655029}.
Since QPUs often have limited connectivity between qubits, circuit transformations have to meet these constraints. To address this, the work \cite{10.1145/3400302.3415621} presents a Monte Carlo Tree Search framework. It allows deeper exploration of the search space. Our compute graph model is in a similar direction, with the main difference that for a limited depth the full space is explored.
The work \cite{Jones2022robustquantum} presents a  method for approximate compilation aimed at optimizing programs for quantum computers by transforming the compilation task into finding the lowest energy state of a quantum system. QUESO is presented in \cite{10.1145/3591254}. It is a tool that automatically synthesizes optimizers specific to different quantum devices. Key elements of QUESO include an algebraic representation of rewrite rules, a method for probabilistically verifying circuit equivalence, a specil data structure called a polynomial identity filter, and an algorithm that applies rewrite rules to optimize circuits. Please note, that the concept of rewrite rules is also applied in this work, but we follow a different approach with our neural guided sampling strategy. The work \cite{10.1145/3519939.3523433} presents Quartz, a framework that explores small circuits to generate potential transformations and verifies them using an automated theorem prover. By applying a cost-based backtracking search with these verified transformations, Quartz optimizes quantum circuits. It can also be summarized as a  rewrite-rule synthesis approach.
The work \cite{10.1145/3341301.3359630} proposes TASO, a computation graph optimizer for deep neural networks. TASO takes as input a list of operator specifications and generates candidate substitutions using the given operators as basic building blocks. All generated substitutions are formally verified against the operator specifications using an automated theorem prover. Transpilation as a pattern matching process has been proposed in \cite{10.1145/3498325}.
A more detailed survery can be found in \cite{yan2025quantumcircuitsynthesiscompilation}. It assesses the viability of an integrated design and optimization framework that includes both logic circuit design and compilation optimization, aiming to minimize manual design efforts and improve execution precision and efficiency, thereby facilitating the hardware implementation and validation of quantum algorithms' advantages. 

 In most publications the core comparisons for evaluating transpiled quantum circuits are done with the software provided by qiskit and frameworks such as Tket, Cirq or BQSKit. We therefore decided to compare our outcome with the most recent transpilers from qiskit and BQSKit.  Whereas qiskit is a well established and easy to use package, supported e.g. by IBM and its online tools, the 
 Berkley Quantum Synthesis Toolkit (BQSKit) is a very powerful software, with
    gate deletion and rule-free gate transpilation algorithms. Several well known algorithms (leap, qsearch, qfast, qpredict, PAS and more) are released as part of BQSKit \cite{younis2022,Schoenberger,Wu21,Davis20,PRXQuantum.2.010324}. As BQSKit outperforms several
commercial compilers (Qiskit, Cirq, Tket), we use it as additional baseline for our experiments, similar to former works. 

\begin{figure}
  \includegraphics[width=0.48\textwidth]{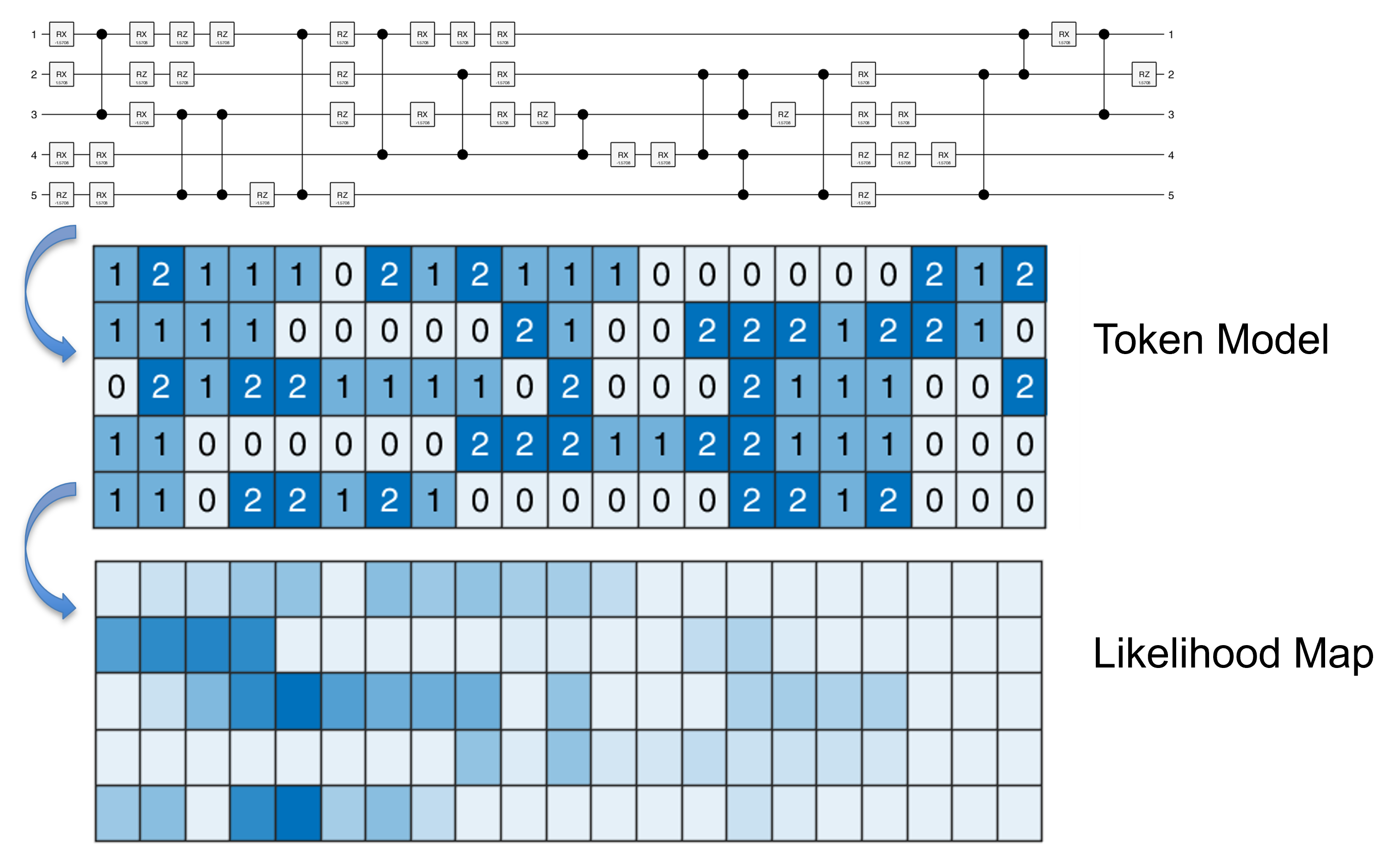}    
\caption{A quantum circuit (top) can be represented as a 2D token model (middle). A neural network is used to predict a likelihood map with areas of sub circuits which are likely reducible. Sampling from this likelihood map leads to an increased sampling efficiency for replacing subcircuits with more efficient ones. 
}
\label{fig:Mot}
\end{figure}

\section{Method}

\subsection{Random search}
From an optimization point of view, the formal problem statement is to find parameters $\theta$ that minimize a loss function $L:\theta \rightarrow \mathbf{R}^{\geq 0}$.
The optimization task  can then be formalized as
\begin{eqnarray}
\theta^\star &\equiv& \underset{\theta \in \Theta}{\mathrm{arg \,min}}\: L(\theta)\\
&=&\{ \theta^\star: L(\theta^\star) \leq L(\theta) ,\ \forall \theta \in \Theta  \}.
\end{eqnarray}
The vector $\theta$ is assumed to be a $p$-dimensional vector of parameters to be optimized.

The simplest method for stochastic optimization is called  \textit{ blind random search} \cite{Spall2003}. Here, a random sample for $\theta$ is generated while not taking into account information about previous samples. 
 The approach can be implemented by drawing a number of examples $\theta$ (typically respecting general boundary properties) and selecting the value of  $\theta$ yielding the lowest $L$ value as an estimate of the optimum. 
 
In general, random search is a non-deterministic algorithm 
in the sense that it produces a reasonable result only with a certain probability. 
Random search algorithms have been proven useful for ill-conditioned global optimization problems, especially for problems where the optimization function can be nonconvex, nondifferentiable or discontinuous over a continuous, discrete, or mixed domain \cite{Blum03,pardalos2002handbook,schoen2002}. 
While blind random search can be a reasonable algorithm for low dimensional $\theta$, it is known that the method can become very slow  for even moderately dimensioned $\theta$. E.g. \cite{Spall2003}
demonstrate as follows: Let $\theta = [0, 1]^p$ be a search space of a $p$-dimensional hypercube with minimum and maximum values of $0$ and $1$ for each component of $\theta$. The goal is to guarantee with probability $0.90$ that each element of  $\theta$ is within $0.04$ units of the optimal value. 
Then, as $p$ increases from one to ten, the number of required loss function evaluations roughly increases by a factor of $10^{10}$.
 This is also known as curse of dimensionality. Thus, even though simple random search strategies can lead to global optimal results (for an infinite time budget), it suffers from sampling efficiency (iterated probing can take too long to find a good solution) and it is not guaranteed that a global optimal solution has been found \cite{7314898}. 

For this reason, many approaches try to exploit assumptions about smoothness of a loss function, sometimes referred to as  localized algorithms \cite{Matyas65}. This leads to a whole family of algorithms, e.g. based on evolutionary algorithms \cite{ZhiHui22}, particle filters (Sequential Monte Carlo methods) \cite{annurev23}, simulated annealing \cite{KOULAMAS199441} or Markov-Chain Monte Carlo Methods \cite{Robert21}.
Unfortunately, even though these approaches can heavily improve sampling efficiency, the manifold of quantum circuits is a heavily non-smooth function, as one gate can heavily change the effect of a quantum circuit.

\begin{figure*}[t] 
\hrule\vspace{-5pt} 
\caption{Compute Graphs}
\label{alg:CG1}
\vspace{-2pt}\hrule\vspace{5pt}

\begin{algorithmic}[1]
\Require Set of elementary quantum gates $\mathcal{OP} = \{O_1, O_2, \ldots\}$, depth $d$
\Ensure Compute graph $\mathcal{G}$ with nodes containing unitary matrices and edges labeled by quantum gates
\State Initialize the graph $(\mathcal{G},\mathcal{G}_E$) with a root node containing the identity matrix $ \mathbb{I}$ and $\mathcal{G}_E= \emptyset$
\State Initialize a queue $\mathcal{Q}$ with the root node $(\mathbb{I}, 0)$, where 0 is the current depth
\While {$\mathcal{Q}$ is not empty}
    \State Dequeue an element $(U, \text{depth})$ from $\mathcal{Q}$
    \If {depth$<d$}
        \ForAll { $O_i \in \mathcal{OP}$}
            \State Compute the new unitary $U_{\text{new}} = O_i \cdot U$
            \If {$U_{\text{new}} \notin$ $\mathcal{G}$}
                \State $\mathcal{G}$=$\mathcal{G}$
                $\cup$ $U_{\text{new}}$
                
                \State $\mathcal{G}_E$=$\mathcal{G}_E \cup (U,U_{\text{new}},O_i) $   
                : Add edge from $U$ to $U_{\text{new}}$ labeled by $O_i$
                \State Enqueue $(U_{\text{new}}, \text{depth} + 1)$ into $\mathcal{Q}$
            \Else
                \State 
                $\mathcal{G}_E$=$\mathcal{G}_E \cup (U,U_{\text{new}},O_i) $  :
                Add an edge from $U$ to the existing node $U_{\text{new}}$ labeled by $O_i$
            \EndIf
        \EndFor
    \EndIf
\EndWhile
\State \Return $\mathcal{G}$
\end{algorithmic}
\vspace{5pt}\hrule
\end{figure*}

\subsection{Optimal unitary factorization for quantum circuit generation}
Given a finite set of available gates, it is possible to systematically explore all potential quantum circuits by constructing a \emph{compute graph} up to a specified depth \cite{rosenhahn2025optimization}. 

The approach begins with the identity operator, denoted by $\mathbb{I}$, as the root node and progressively constructs quantum circuits by selecting gates from a predetermined set of elementary quantum gates, $\mathcal{OP} = \{O_1, O_2, \ldots\}$. The resulting compute graph is a graphical representation where nodes encapsulate unitary matrices, and edges symbolize elementary unitary operations $O_i \in \mathcal{OP}$.

Initially, the graph is established with $\mathbb{I}$ serving as the root node. A gate $O_i$ is chosen and applied to this root, yielding a new node through the multiplication of the selected gate with the unitary matrix of the parent node. If the resulting unitary matrix already exists as a node within the graph, a direct edge is created from the parent node to the existing node. Otherwise, a new node is generated and linked to the parent node. It is important to note that unitary matrices are compared using a numerical tolerance of $10^{-5}$, and any global phase differences are compensated.

As the graph develops, the unitary matrices formed become nodes. The quantum circuit corresponding to a target unitary can be found by identifying the shortest path from the root node to the target node and compiling the operators associated with the edges along this path. Figure \ref{alg:CG1} summarizes the pseudocode of the algorithm.

The compute Graph is used to generate a dataset $\mathcal{D}=\{(U_i; [ o^i_1 \ldots  o^i_m] )\}$. This dataset consists of the node unitaries $U_i$ and its shortest realization as quantum circuit. This optimal realization is found by collecting the edges of the shortest path $o^i_1 \ldots  o^i_m$ in the compute graph up to the node of the unitary $U_i$.

Please note, that this compute graph is only feasible to compute up to a certain depth and with a limited amount of available gates, as the combinatorial space is exploding exponentially. Therefore, it is specifically useful for settings with a limited amount of gates which is naturally the case for hardware realizable gates and the transpilation task. As the amount of gates is discrete, it also requires that the angles are discretized. 

For this reason, a local term replacement scheme is proposed to iteratively reduce a large quantum circuit, as outlined in the next subsection. This local term replacement can also be read as local resynthesis.

\begin{figure*}[t] 
\hrule\vspace{-5pt} 
\caption{1D Circuit Optimization}
\label{alg:1DOpt}
\vspace{-2pt}\hrule\vspace{5pt}
\begin{algorithmic}[1]
\Require Token chain $U = O(L) O(L-1) \dots O(1)$, tolerance $\epsilon = 10^{-5}$
\Require Dataset $\mathcal{D}=\{(U_i; [ o^i_1 \ldots  o^i_m] )\}$ of unitaries and their optimal decomposition
\Ensure Optimized token chain $U$
\State \textbf{function} RandomSearch($U$)
    \While {stopping criteria not met}
        \State Randomly select a position $m$ and length $n$ such that $m+n-1 \leq L$
        \State Extract the subset $U_s = O(m+n-1) \dots O(m)$
        \If {$U_s \in \mathcal{D} $ (up to $\epsilon$)}
        \State $(O^r(p) \dots O^r(1)) = \mathcal{D}(U_s)$ : Replace token chain of $U_s$ with optimal token chain
        \State Update $U = O(L) \dots (O^r(p) \dots O^r(1)) \dots O(1)$
        \EndIf
        \State Randomly exchange commutative blocks
    \EndWhile
    \State \Return  $U$
\end{algorithmic}
\vspace{5pt}\hrule
\end{figure*}

\subsection{Quantum Circuits as token chains}
\label{Sec:QCToken}
To realize a quantum circuit as a 1D token chain, our method requires a discrete set of available gates. For example this can be the gates $RX$, $RZ$ and $CZ$ for a NISQ architecture. Now, each possible operation is mapped to a unique number (the token). In the following, it is described  for the simple case of three qubits.
 Each gate acts on either one (e.g. a $RX$ gate) or two  qubits (e.g. a $CZ$ gate ). The 1-qubit gates take additionally a parameter $\theta$. Four discretized angles $[-\frac{\pi}{2},-\frac{\pi}{4}, \frac{\pi}{4},\frac{\pi}{2}]$ are taken for this example. For three qubits $[1, 2, 3]$  it is now possible to generate 12 gates, namely, 
 \begin{align*}
     &\left[ RX(1, -\tfrac{\pi}{2}), \ldots, RX(1, \tfrac{\pi}{2}),\right. \\ 
     &\quad RX(2, -\tfrac{\pi}{2}), \ldots, RX(2, \tfrac{\pi}{2}), \\
      &\quad\left.RX(3, -\tfrac{\pi}{2}), \ldots, RX(3, \tfrac{\pi}{2})\right],
 \end{align*} 
 which are the tokens $1 \ldots 12$.
The unique tokens $13 \ldots 24$ can be used to represent
\begin{align*}
    &\left[RZ(1, -\tfrac{\pi}{2}), \ldots, RZ(1, \tfrac{\pi}{2}),\right. \\
    &\quad RZ(2, -\tfrac{\pi}{2}), \ldots, RZ(2, \tfrac{\pi}{2}), \\
    &\quad\left.RZ(3, -\tfrac{\pi}{2}), \ldots, RZ(3, \tfrac{\pi}{2})\right].
\end{align*} 
Finally, it is possible to combine the $CZ$ gates to \begin{align*}
    \left[CZ(1,2),CZ(1,3),CZ(2,3)\right],
\end{align*} 
for the tokens 25 $\ldots$ 27.
Please note, that the $CZ$-gate is commutative across the qubits. Therefore, there is no need to generate e.g. $CZ(2,1)$. This gate set results for three qubits and the provided angular discretization in an operator pool of 27 gates.
The quantum circuit $RX(1, -\frac{\pi}{2}) CZ(1,2) RZ(3, -\frac{\pi}{2}) $ can then be expressed as the token chain ${\tt [O(1) O(25) O(21)\ }$. 
Each token can be encoded as a vector of the form [operatorID, controlqubit, targetqubit, angle]. In our experiments the operator ID is encoded using a binary encoding. E.g. for our three gates, $RX$ is encoded as $(1,0,0)$,  $RZ$ as $(0,1,0)$ and $CZ$ as $(0,0,1)$. The quantum gate $CZ(1,2)$, or $RZ(2,\frac{\pi}{2})$ can therefore be represented as 6D vector of the form 
\begin{eqnarray*}
(\underbrace{0,0,1}_{CZ},\underbrace{1}_{start qubit},\underbrace{2}_{target qubit},\underbrace{0}_{angle})&&\\
(\underbrace{0,1,0}_{RZ},\underbrace{2}_{start qubit},\underbrace{0}_{target qubit},\underbrace{\frac{\pi}{2}}_{angle})&&
\end{eqnarray*}
\subsection{1D Random Search Based Quantum Circuit Optimization}
Random search has been used in the past  
for quantum circuit optimization \cite{rosenhahn2025optimization}:
A quantum circuit can be viewed as a sequence of tokens, such as $  U = O(L)O(L-1)\cdots O(1) $ = {\tt O(L) O(L-1) \dots O(1)}, where each token represents an operator within the quantum circuit.

Motivated from formal languages, each token corresponds to an element of an alphabet and a series of tokens can be seen as a word as an element of a formal language. Formal grammars consist of a set of so-called production and rewriting rules for transforming words. Each rule specifies a replacement (or local resynthesis) of a particular token-chain  with another. Several books cover this fundamental topic, e.g.
  \cite{Leuuwen91}.
The random search and replace algorithm for quantum circuit optimization works as follows: Given a token chain $ U= O(L) O(L-1) \dots O(1) $, first, a random connected subset, e.g. at position $m$ of length $n$ is selected, 
\begin{eqnarray*}
U &=& O(L) \dots \underbrace {(O(m+n-1) \dots O(m))}_{\text{length n}} \dots  O(1)\\
 &=& O(L) \dots (U_s) \dots  O(1).
\end{eqnarray*}
The subset $ O(m+n-1) \dots O(m) $ generates a unitary  $U_s$.
In \cite{rosenhahn2025optimization} this unitary is compared to a database of unitaries where their optimal gate decompositions are known. This database has been generated by using a full compute graph up to a certain depth \cite{RosOsb2023a}, as summarized in the earlier subsection. If the
unitary  $U_s$ is found in the database (up to an error $\epsilon$ and the known optimal factorization is shorter than the selected set, it can be replaced with a chain of alternative gates $ U_p = O^r(p) \dots O^r(1)$. 
\begin{eqnarray*}
U &=& O(L) \dots \underbrace {(O(m+n-1) \dots O(m))}_{\text{length n}} \dots  O(1) \\
&=&  O(L) \dots (U_p) \dots  O(1) \\
&=& O(L) \dots \underbrace {(O^r(p) \dots O^r(1))}_{\text{length p}} \dots  O(1).
\end{eqnarray*}
Since many blocks in a quantum circuit are commutative (e.g. when acting on different qubits), such blocks can be randomly exchanged and the process is iterated. Figure \ref{alg:1DOpt} summarizes the algorithm as pseudocode.

We observed that optimizing a 1D token chain can be highly sampling inefficient due to the high amount of commutative elements in there and due to the random nature in searching for more efficient circuit blocks.
For this reason, in \cite{rosenhahn2025optimization} a random forest has been proposed to reject samples  early which are likely non reducible. Even though this process is fast, there are still many unnecessary proposals generated and rejected. 

\begin{figure}
\includegraphics[width=0.45\textwidth]{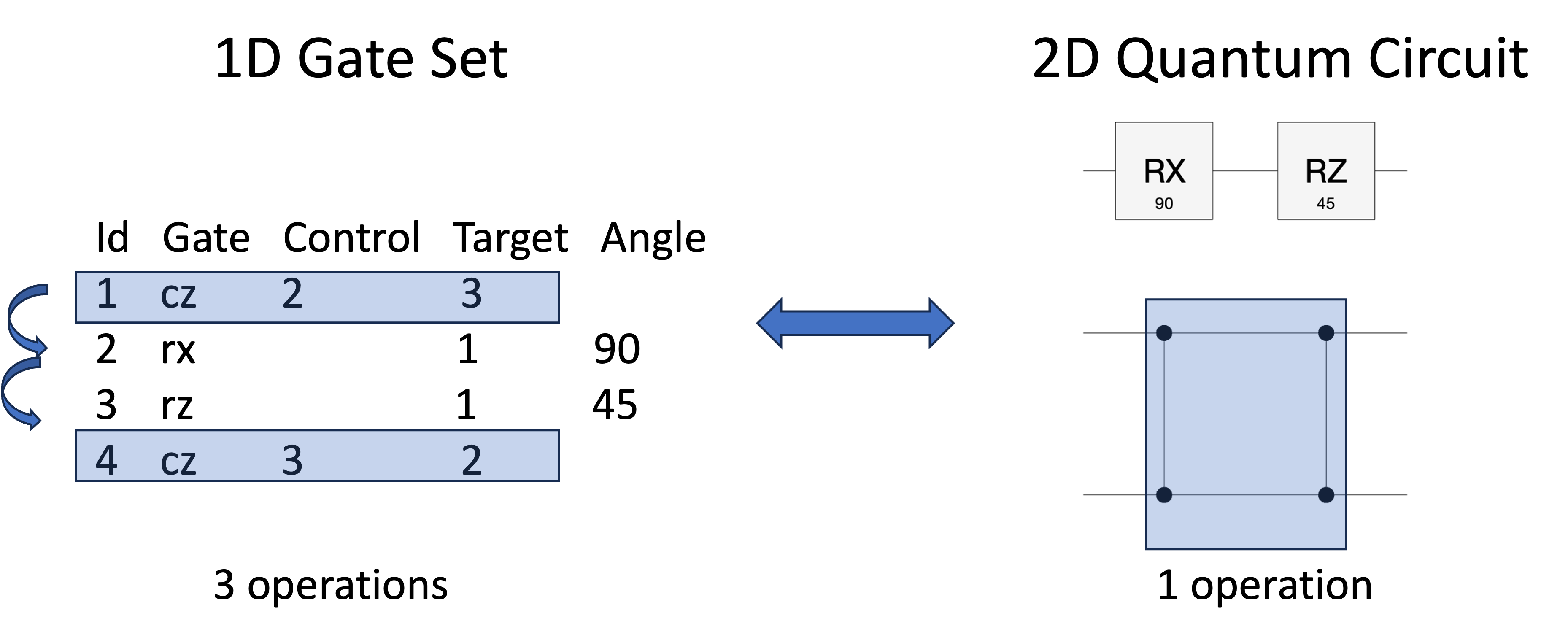}    
\caption{Left: The example quantum circuit consisting of 4 consecutive gates. To reduce this sequential gate set, 3 operations are required, two operations to shuffle the CZ-gates one after another. Afterwards follows the reduction of the two CZ-gates which form an identity matrix. Right: The 2D representation of the quantum circuit leads to a structure which allows to immediately identify that the rotations have no impact on the CZ gates and the block can be directly removed.
}
\label{fig:1Dvs2D}
\end{figure} 
Figure \ref{fig:1Dvs2D} visualizes the inefficiencies when working on the 1D gate set forming a quantum circuit: 
The left part shows a quantum circuit represented by  4 consecutive gates. To reduce this circuit, 3 operations are required, two operations to shuffle the CZ-gates one after another. This is possible since the rotation gates act on different qubits and are therefore commutative in the circuit. After this, it is possible to reduce the two neighboring CZ-gates which form an identity matrix. The right part of the figure shows the 2D representation along the qubits. This representation leads to a structure which allows to immediately identify that the rotations have no impact on the CZ gates and the CZ block can be directly removed.
We therefore propose neural guided sampling:
a neural network is used to determine which parts in this 2D representation are likely reducible by predicting a likelihood map. This map can then be used for a more efficient sampling. 

\begin{figure}
  \includegraphics[width=0.45\textwidth]{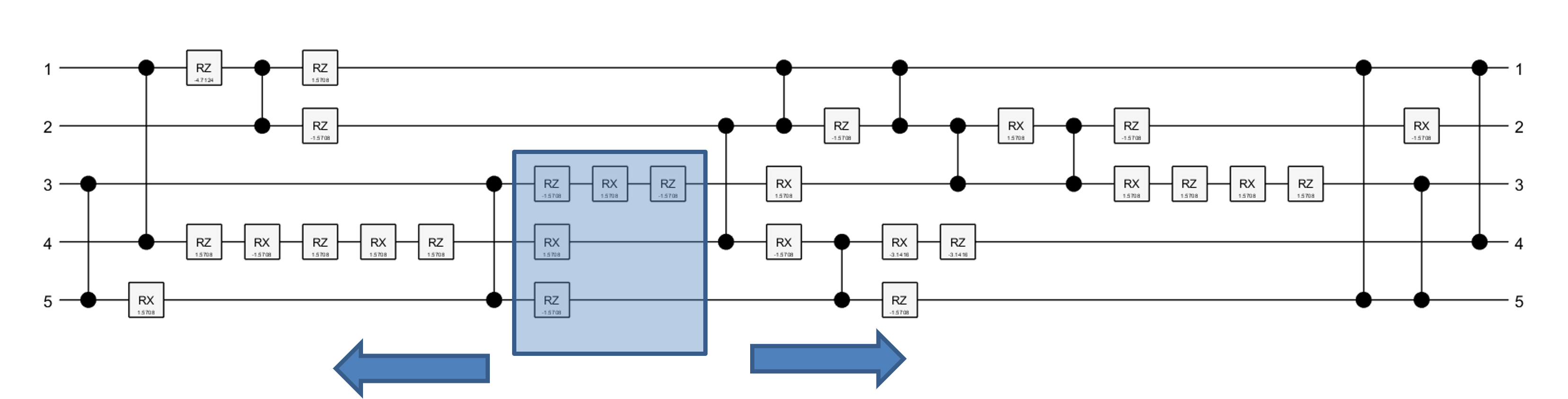}    
\caption{After selecting a reducible quantum circuit fragment, the quantum circuit is separated into three segments, such that the middle circuit block can be replaced with a more efficient one. This local resynthesis is based on a database lookup with 
unitaries and their optimal decomposition in base gates. 
}
\label{fig:Sep}
\end{figure} 

\subsection{2D Random and Neural Guided Sampling Based Quantum Circuit Optimization}
\label{SecNGS}
As mentioned in the previous subsection, using a 1D token chain representation for a quantum circuit can be inefficient due to the high amount of commutative elements which makes it hard to look up the reducible circuit elements. We therefore propose two major contributions in this work, (a) to sample from a 2D token representation and (b) to make use of neural guided sampling to efficiently reduce the quantum circuit length.
Here, neural guided sampling means to 
use a neural network for computing a likelihood map. Sampling from this map should prefer
quantum circuit areas, which are likely reducible.
This concept has previously been applied in the field of RANSAC: The Random Sample Consensus (RANSAC) is an iterative technique used to optimize the parameters of a mathematical model or function based on a set of observational data that includes outliers \cite{Fischler81}.
RANSAC has been used in the past  for geometric model fitting, e.g. homography estimation, fundamental matrix estimation, structure from motion, image stitching and more \cite{Torr1997,torr2000a,1498749,6330965}. There are also extensions of multi-model fitting which contain the task to fit multiple models simultaneously, e.g. to estimate different fundamental matrices for independently moving objects \cite{9010674}. As these approaches also suffer from sampling efficiency, 
neural networks have been successfully used to guide this process \cite{KluBra2020,KlugerTPAMI2024} and we adopt this strategy to the domain of quantum circuit reduction. 

Figure \ref{fig:Mot} shows the general idea:
A quantum circuit can be represented as a 2D token model, similar to a small 2D image. This 2D array carries as third dimension the gate representation (e.g. 3D vector) explained in Section \ref{Sec:QCToken}. This 3D token representation is the input of the neural network. 
Based on this token representation, a neural network predicts the likelihoodmap with areas, which are likely reducible. See
e.g. Figure \ref{fig:Mot}, where darker colors indicate areas which are more likely reducible than brighter ones. 
We then sample circuit parts based on this likelihood map \cite{george1993variable}.

\begin{figure*}[t] 
\hrule\vspace{-5pt} 
\caption{2D Neural Guided Sampling}
\label{alg:floating_version}
\vspace{-2pt}\hrule\vspace{5pt}
 \begin{algorithmic}[1]
        \Require $G$: Gate set 
        \Require $qc$: Quantum circuit 
        \Require $NN$: Neural Net for Neural Guided Sampling 
        \Procedure{optimizeCode}{$qc$, $G$, $NN$}
            \Repeat
                \State $[LCode, MCode, RCode] \gets \texttt{FindNGSSubGrid}(qc, NN)$ 
                \State $NewCode \gets \texttt{reduceCode}(MCode)$
              \State $qc \gets \texttt{concatenate}(LCode, NewCode, RCode)$
            \Until{termination condition is met}
       \State  \Return $qc$
        \EndProcedure
    \end{algorithmic}
\vspace{5pt}\hrule
\end{figure*}

\begin{figure}
  \includegraphics[width=0.48\textwidth]{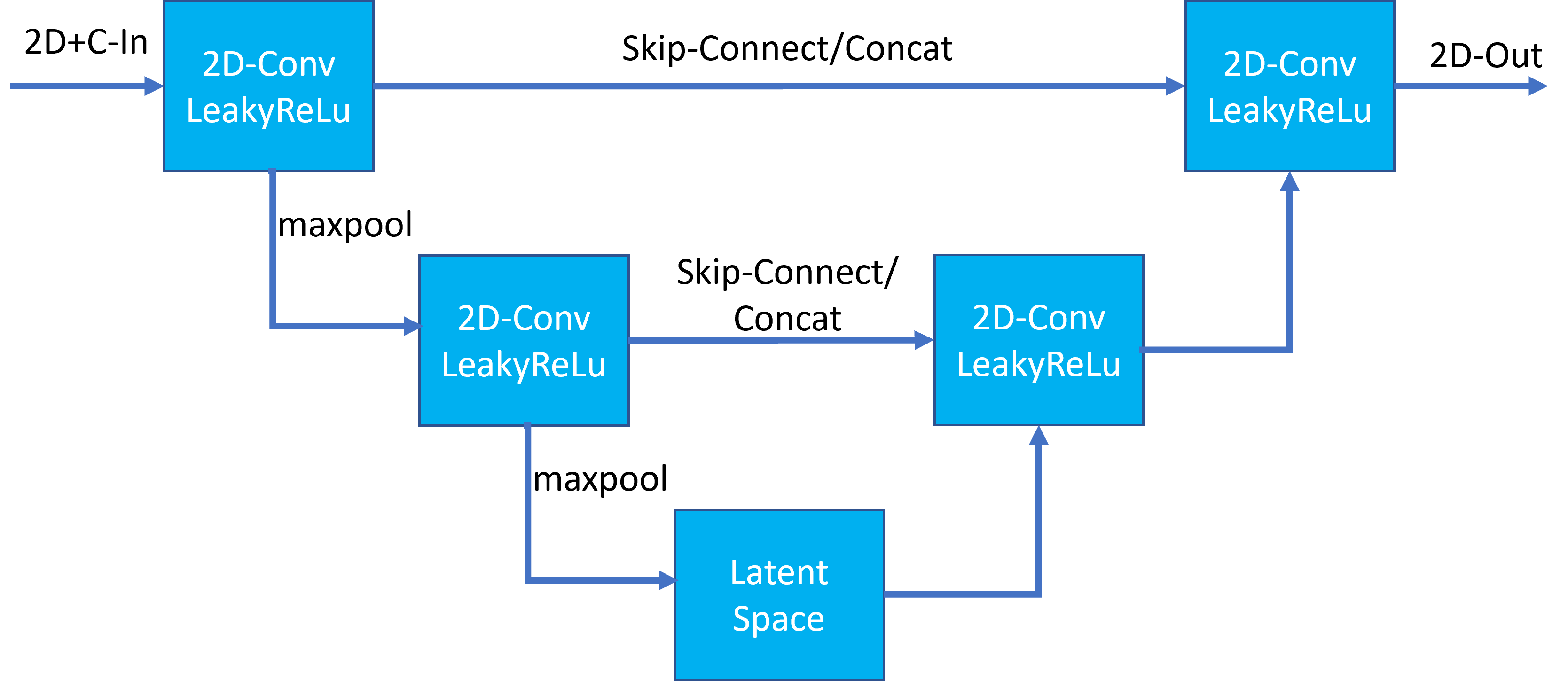}  
\caption{General structure of the Encoder-Decoder architecture for prediciting the likelihood map which is afterwards used for neural guided sampling.  
}
\label{fig:EncDecNet}
\end{figure}

For training the transformation from the input 2D token chain to the likelihood map, we use a typical encoder-decoder architecture, see also Figure~\ref{fig:EncDecNet} for a summary. It consists of two main components: an encoder that processes the input sequence (the representation of the quantum circuit) and converts it into a fixed-length vector representation (the latent space). Afterwards, a decoder generates an output sequence based on this encoded representation. Autoencoders \cite{doi:10.1126/science.1127647}, Variational Autoencoder \cite{MAL-056} or UNets \cite{UnetRef} are typical examples for encoder-decoder based neural networks. 
For our experiments we decided on a UNet like architecture with 2 blocks for downsampling and 2 blocks for upsampling, as well as skip connections. 
All convolution blocks consist of $3 \times 3$ convolutions along $16,16,32$ channels during down and upsampling. Each block consists of two successive convolutions followed by a leaky relu function and a  $2 \times 2$ max-pooling layer.
 Furthermore, we applied dropout layers \cite{Srivastava14} to reduce the effect of overfitting. Further hyper-parameters we used are a mini batch size of 20, a learning rate of 0.002, epoch shuffling and we used an adam optimizer. As input to the network we encoded the quantum circuit as a 2D image with the amount of qubits in one dimension and the depth of the circuit as second dimension. Each qubit is encoded with its properties along a channel dimension. A binary flag encodes the operator type (e.g. $RX$, $RZ$ or $CZ$) and afterwards we stack the control qubit, target qubit, as well as the angle, see Section \ref{Sec:QCToken}. If one of these properties is unused, the value is set to zero. Thus, for our experiments, we arrived at 8-10 channels for each gate. 
 
 After selecting a reducible quantum circuit fragment, see Figure \ref{fig:Sep}, the quantum circuit is separated into three parts and the algorithm seeks for a more efficient representation of the middle (the selected) block. 
 The middle circuit  consists usually of many wires that are not connected to these gates and they only cause an increase of the underlying dimensions. Thus, we drop the non-needed wires and map the selected gates to a subspace only containing required qubits. E.g. in Figure \ref{fig:Sep}, the first two qubits can be ignored \cite{rosenhahn2025optimization} and instead of optimizing a $2^6$-dimensional unitary, we only have to optimize for a $2^4$-dimensional unitary matrix, representing the circuit.
 To compare the original subblock unitary with the replaced unitary, we numerically accept a tolerance of $10^{-5}$ and we compensate for a global phase. 
 For optimization, we rely on an optimal factorization of a unitary using a given compute graph, as presented in  \cite{rosenhahn2025optimization}.
 After the analysis in the smaller qubit space and potential quantum circuit
 reduction, the resulting circuit block is mapped back to the original size. Thus, the middle circuit block can afterwards be replaced by a more efficient one, see Figure \ref{fig:Sep}. This process of identification of possible reducible sub-circuits is iterated until a fixed time budget or target length is reached.

\section{Experiments}
We decided to perform experiments on two gate sets. The first gate set we use is common for ion-trap architectures and comprises $RX,RY,RZ$  and $RXX$ gates. The
second gate set we use is common for NISQ-architectures and comprises $RX, RZ$ and $CZ$ gates.
Furthermore, we solely evaluate the performance on randomly generated quantum circuits. 
 It ensures that our approach is not influenced by a hidden bias. Thus, the data samples are maximally diverse. Furthermore, it allows to  verify the generalization capacities of the neural networks' predictive power.
\subsection{Generating Training Data}
For the experiments on both architectures there is at first a need to generate sufficient training data to optimize the used neural networks. Therefore we generate/sample  random circuits for each architecture separately.
The target likelihood map is initialized as an array of zero entries
and we use random sampling based on an equal distribution to find sub-blocks which are reducible. Once such a block has been found, the target likelihood map is incremented at these locations. This process is repeated until a time budget has been used up (we used 10 min for each example). Then the array is normalized to define the target likelihood map the neural network is proposed to predict from the input circuit.
 It is  like a heat map according to Figure \ref{fig:Mot}. This heatmap is the training target of the neural network.
For our experiments we generated a dataset with $10,000$ examples for a respective input (quantum circuit)/output (likelihood) pair. This has been done twice for both gate sets we have been using. This dataset is then used for training the neural network. 
\begin{figure*}
  \includegraphics[width=0.95\textwidth]{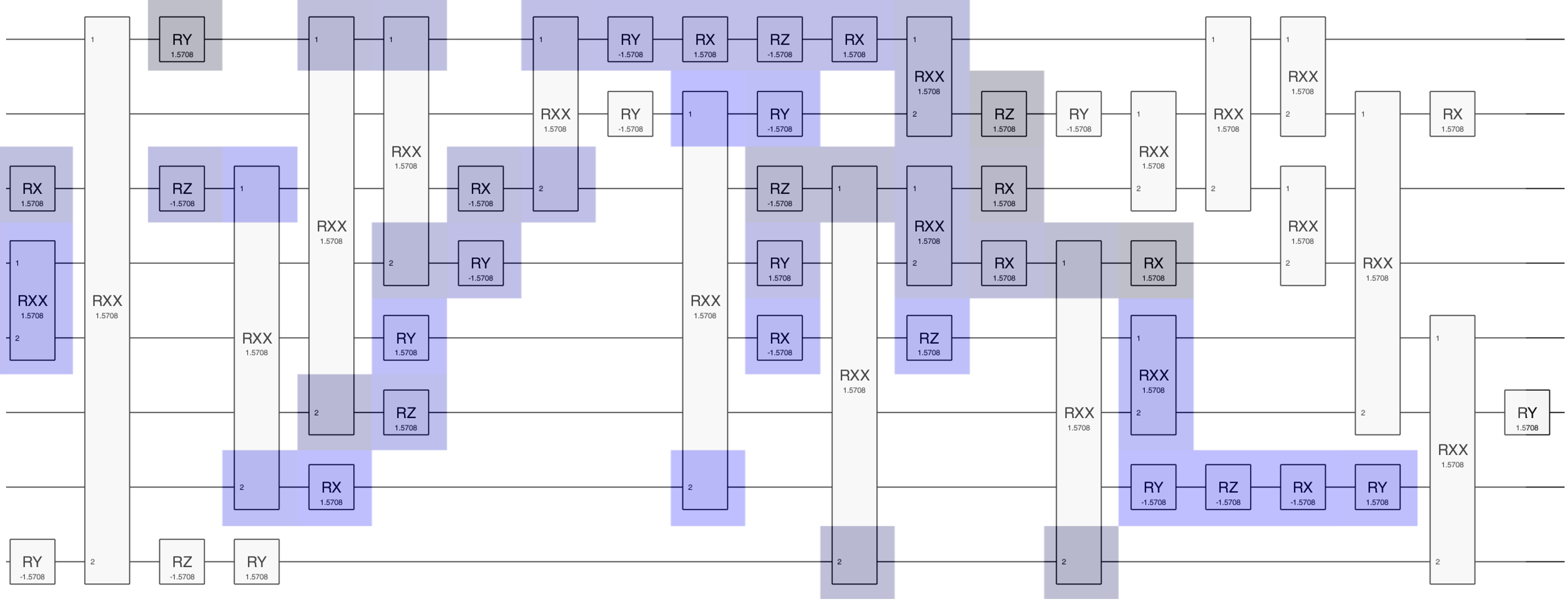}  
\caption{Overlay of a random generated quantum circuit with an underlined likelihood map for quantum circuit reduction opportunities. Darker colors indicate parts of the quantum circuit which are likely reducible.
}
\label{fig:Overlay}
\end{figure*} 

Figure \ref{fig:Overlay} visualizes a training example as overlay of the random generated quantum circuit (as input) and the underlined likelihood map for quantum circuit reduction opportunities as (desired) output of the neural network. A darker color indicates a higher likelihood for being able to replace a local gate block with a more efficient gate sequence. 

\begin{figure}
  \includegraphics[width=0.45\textwidth]{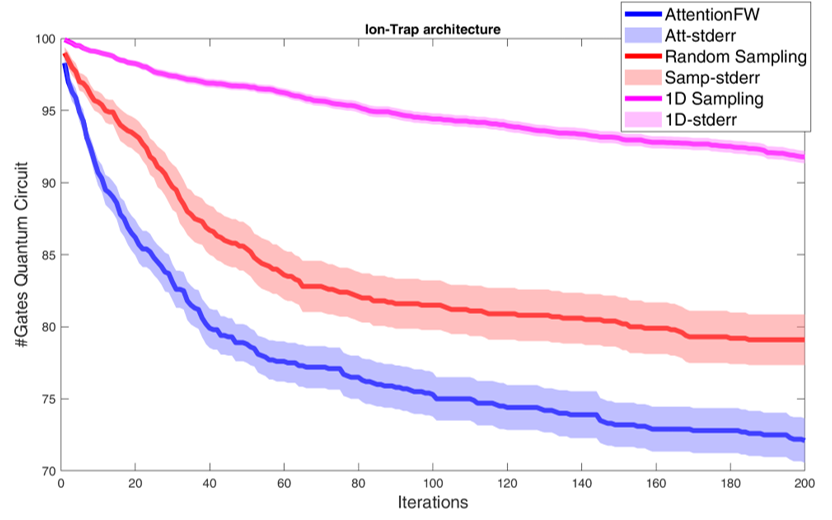}  
\caption{Neural Guided Sampling in comparison to random sampling for an Ion Trap architecture.
}
\label{fig:ConvIon}
\end{figure} 

\begin{figure}
  \includegraphics[width=0.45\textwidth]{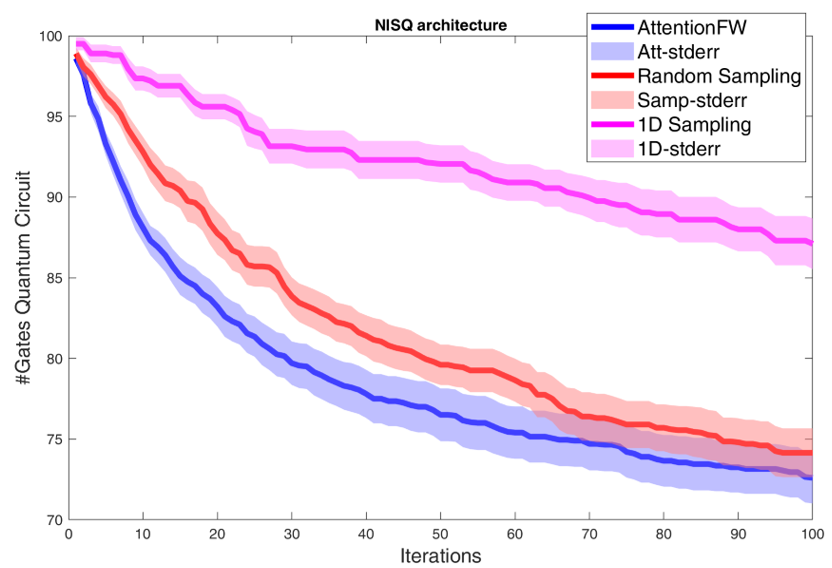}  
\caption{Neural Guided Sampling in comparison to random sampling for a NISQ architecture.
}
\label{fig:ConvNisq}
\end{figure}

\subsection{Evaluation}
The main competitor of our approach is the 1D variant \cite{rosenhahn2025optimization} and we will show in our experiments that the 2D neural guided sampling method achieves (a) more efficient quantum circuits (code with less gates to represent the same unitary) and (b) the generation of the circuits is much faster. We further compare the results of our method with qiskit \cite{QiskitTextbook:2020} and
BQSKit \cite{BQSKIT} on different optimization levels.

Figures  \ref{fig:ConvIon}  and \ref{fig:ConvNisq}  summarize our first experiment. In this experiment we sample 100 random quantum circuits with a size of 100 gates. Then we run a fixed amount of iterations for the 1D approach presented in \cite{rosenhahn2025optimization}, a 2D random sampling based approach (using an equal distribution) and the neural guided sampling.
The goal is to demonstrate that neural guided sampling is beneficial over a 1D and naive 2D random sampling. 
 As the local resynthesis scheme works iteratively, we show in Figures  \ref{fig:ConvIon} and \ref{fig:ConvNisq}  on the $x$-axis the amount of iterations of the algorithm and on the $y$-axis the gate count. Thus, with an increasing amount of iterations, we expect that the algorithms are able to continuously reduce the quantum circuit input size, leading to a monotonic decreasing function. Both figures show that over these sampled circuits the proposed neural guided sampling based approach requires less iterations for generating more efficient quantum circuits, compared to the 1D competitor \cite{rosenhahn2025optimization} and a naive 2D sampling scheme.
\begin{figure}
  \includegraphics[width=0.45\textwidth]{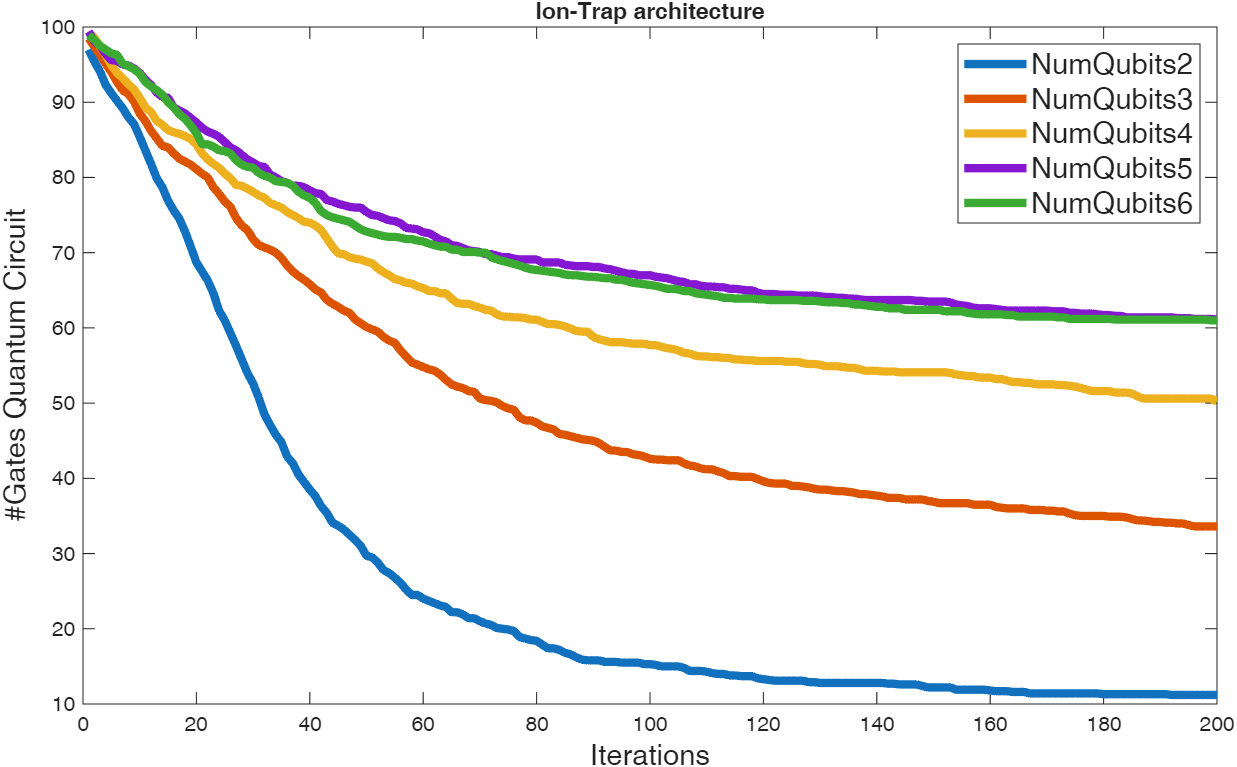}  
\caption{Convergence of our proposed Neural Guided Sampling for varying amount of qubits for an ion-trap architecture (mean over 20 runs).
}
\label{fig:ConvNQub}
\end{figure} 

\begin{figure}
  \includegraphics[width=0.45\textwidth]{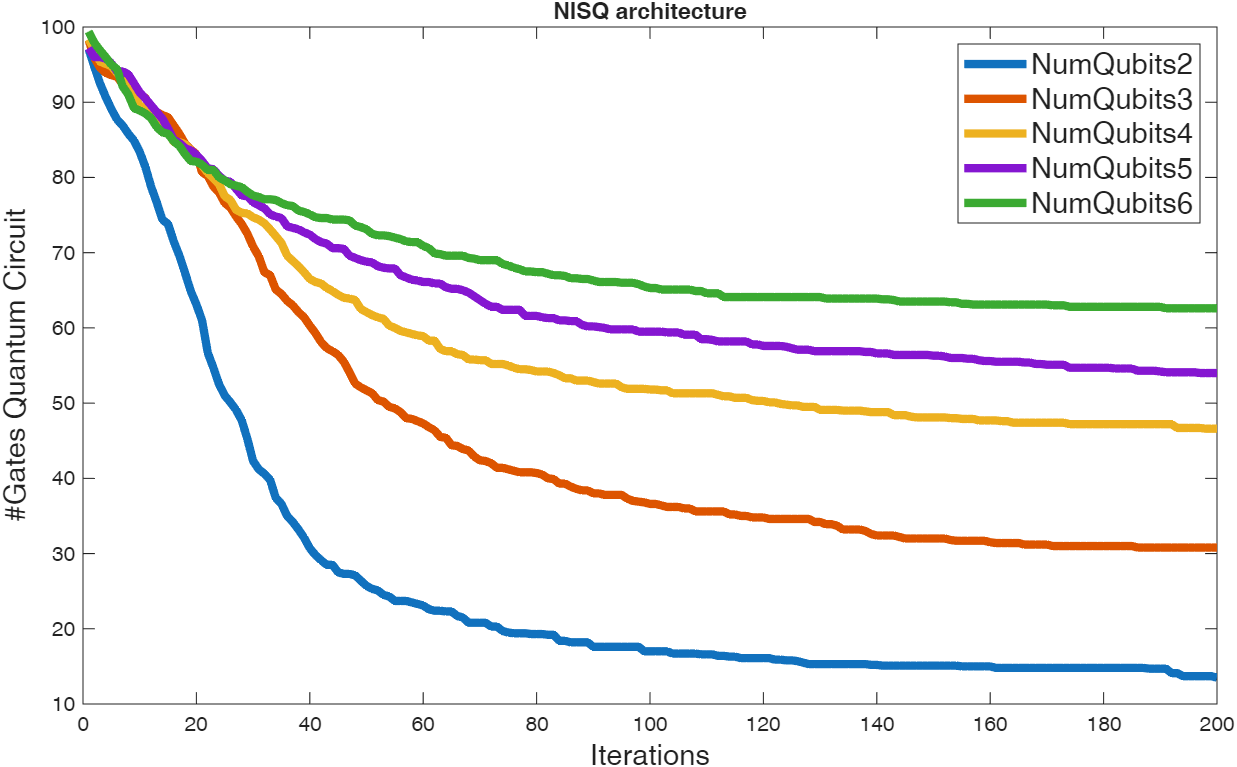}  
\caption{Convergence of our proposed Neural Guided Sampling for varying amount of qubits for a NISQ architecture (mean over 20 runs).
}
\label{fig:ConvNQubNI}
\end{figure} 

Figures \ref{fig:ConvNQub} and \ref{fig:ConvNQubNI} show the convergence of our proposed Neural Guided Sampling for varying amounts of qubits for both, the ion-trap architecture and the NISQ architecture. Similar to 
Figures  \ref{fig:ConvIon} and \ref{fig:ConvNisq}, 
we sample 20 random quantum circuits with a size of 100 gates. Then we run a fixed amount of iterations and plot the mean gate count over the iterations when varying the circuit from 2 to 6 qubits. 
As a higher number of qubits leads to highly sparse configurations, it is getting harder to identify reducible or optimizable sub-blocks in a multi-qubit system.  Therefore, it is expected that the algorithm performs better and faster for the smaller qubit regime.

 In the following, we demonstrate on several experiments that our approach is not only sample efficient, but yields significantly shorter quantum circuits compared to the current state of the art.

\begin{table}
{\footnotesize
\begin{tabular}{|c|c|c|c|c|c|}
\hline
Method & $\#$ rx $\downarrow$ &   $\#$ ry $\downarrow$& $\#$ rz $\downarrow$ & $\#$ rxx $\downarrow$ &TTB $\downarrow$\\
&  &   &  &  & (sec) \\
\hline
original & 27 & 29 & 36 & 58 &-\\
\hline
Q-L1 & 21 & 23 & 28 & 58&- \\
Q-L2 & 21 & 23  & 28 & 58&-  \\
Q-L3 & 21 & 23 & 28 & 58&- \\
\hline
B-L2 & 41 & 9 &  47 & 51 & - \\
B-L3 & 34 &  4 &  47 &   47 & - \\
B-L4 & 55 & {\bf 0 }&  60 &   47 & - \\
\hline
1D \cite{rosenhahn2025optimization} & 13 & 21 & 32 & 54 &120 \\
& & & & & { ($\pm$30)}  \\
\hline
{\bf ours} & {\bf 6 } &  14 & {\bf 24} & {\bf 46} & {\bf 4.5 }\\
& & & & & { ($\pm$1.5)}\\
\hline
\end{tabular}
}
\caption{Quantum circuit optimization example for an ion-trap architecture and comparison of our method to the  qiskit optimizer on levels 1-3, the BQSKit optimizer on levels 2-4 and the 1D token model presented in \cite{rosenhahn2025optimization} (8 qubits).
TTB stands for \textit{time to beat} and indicates the time which is required be be better than qiskit lvl 3. Compared to the 1D token model, the proposed 2D Neural Guided sampling is much faster and provides better results.
}
\label{tab:IOComp}
\end{table}

\begin{table}
{\footnotesize
\begin{tabular}{|c|c|c|c|c|}
\hline
Method & $\#$ rx $\downarrow$ &  $\#$ rz $\downarrow$ & $\#$ cz $\downarrow$ &TTB (sec) $\downarrow$\\
\hline
original     & 42 & 43 & 65 &-\\
\hline
Q-L1 & 28 & 27  & 61&- \\
Q-L2 & 26 & 16   & 41&-  \\
Q-L3 & 26 & 16  & 41&- \\
\hline
B-L2 & 29 & 27 &  64 & - \\
B-L3 & 36 & 38 &  51 & - \\
 B-L4 & 65 & 47 &  47 & - \\
 \hline
 1D \cite{rosenhahn2025optimization} & 23 & 13 &  37 &120 \\
 & & & & { ($\pm$50)} \\
\hline
{\bf ours} & {\bf 22 } & {\bf 10} & {\bf 36} & {\bf  20.0} \\
 & & & & { ($\pm$5)} \\
\hline
\end{tabular}
}
\caption{Quantum circuit optimization example for a NISQ architecture and comparison of our method to the  qiskit optimizer on levels 1-3, the BQSKit optimizer on levels 2-4 and the 1D token model presented in \cite{rosenhahn2025optimization} (8 qubits).
TTB stands for \textit{time to beat} and indicates the time which is required be be better than qiskit lvl 3. Compared to the 1D token model, the proposed 2D Neural Guided sampling is much faster and provides better results.
}
\label{tab:NIComp}
\end{table}

Tables \ref{tab:IOComp} and \ref{tab:NIComp}  
summarize results for the transpilation of a random example circuit for a NISQ and an ion-trap based architecture. Each column shows the gate count of the used gates in the quantum circuit.
The symbol $\downarrow$ indicates that the optimization target is to reduce the gate count.
Our method ({\bf ours}) is compared to the  qiskit optimizer on levels 1-3, the BQSKit optimizer on levels 2-3 and the 1D token model presented in \cite{rosenhahn2025optimization}. The column TTB contains the  \textit{time to beat} and indicates the time which is required to be better than qiskit lvl 3. Whereas the 1D token model requires on average 2 minutes, our method is already better within several seconds. The variance is estimated from 10 runs.
\begin{figure}[ht]
\centering
\includegraphics[width=0.475\textwidth]{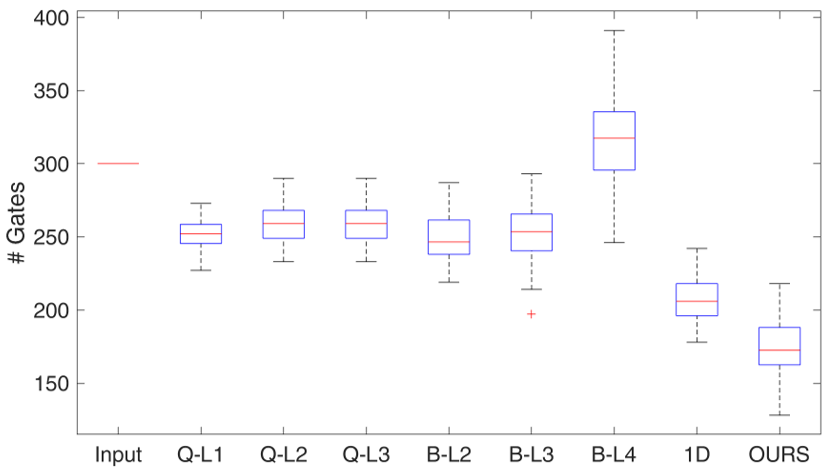}    
\caption{
Statistical summary of the baseline compilation methods qiskit and BQSKit, compared to our proposed method.
100 different randomly sampled
quantum circuits of length 300 (8 qubits) are reduced using the optimization levels 1-3 in qiskit (denoted as Q-L1 ... Q-L3) and the optimization levels 2-4 in BQSKit (denoted as 
B-L2 ... B-L4). 1D denotes the outcome of \cite{rosenhahn2025optimization}.
Our results are denoted as OURS.
 The y-axis shows a box-plot. As the Input is of size 300, there is no standard deviation in this column.
The gate set consists of $RX$, $RY$, $RZ$ and $RXX$ gates, as example for an ion trap architecture.
}
\label{fig:PerfStatIon}
\end{figure}

\begin{table}
\begin{tabular}{|c|c|c|c|c|}
\hline
Method & $\#$ RX $\downarrow$ &   $\#$ RY $\downarrow$& $\#$ RZ $\downarrow$ & $\#$ RXX $\downarrow$ \\
\hline
In &  63 ($\pm$ 7) &  64 ($\pm$ 7)&  63 ($\pm$ 7) &  110 ($\pm$ 8)\\
Q-L1 &  42 ($\pm$ 6) &  47 ($\pm$ 6)&  52 ($\pm$ 6) &  109 ($\pm$ 8) \\
Q-L2 &  45 ($\pm$ 6) &  49 ($\pm$ 5)&  56 ($\pm$ 9) &  109 ($\pm$ 8)  \\
Q-L3 &  45 ($\pm$ 6) &  49 ($\pm$ 5)&  56 ($\pm$ 9) &  109 ($\pm$ 8) \\
\hline
B-L2 & 65 ($\pm$ 7) &  17 ($\pm$ 6)&  77 ($\pm$ 9) &  90 ($\pm$ 8)  \\
B-L3 &  69 ($\pm$ 8) &  12 ($\pm$ 6)&  85 ($\pm$ 10) &  87 ($\pm$ 8)  \\
B-L4 & 105 ($\pm$ 7) &  0 ($\pm$ 0)&  131 ($\pm$ 14) &  84 ($\pm$ 9)  \\
\hline
1D \cite{rosenhahn2025optimization}& 22 ($\pm$ 4) &  42 ($\pm$ 6)&  43 ($\pm$ 6) &  100 ($\pm$ 9)\\
{\bf ours} & 14 ($\pm$ 3) &  36 ($\pm$ 6)&  37 ($\pm$ 6) &  86 ($\pm$ 9)\\
\hline
\end{tabular}
\caption{
Statistics for Quantum circuit optimization  for an ion trap architecture and comparison of our method to the  qiskit optimizer on levels 1-3 and the BQSKit compiler on optimization levels 2-4 over 100 runs
(8 qubits). The values are the mean and in brackets we provide the standard deviation.
}
\label{tab:IOCompL1}
\end{table}

\begin{table}
\begin{tabular}{|c|c|c|c|}
\hline
Method & $\#$ RX $\downarrow$ &   $\#$ RZ $\downarrow$ & $\#$ CZ $\downarrow$\\
\hline
In & 80 ($\pm$ 8) & 79 ($\pm$ 7)  &  141 ($\pm$ 9)  \\
Q-L1 & 58 ($\pm$ 6) & 60 ($\pm$ 6)  &  129 ($\pm$ 9)  \\
Q-L2 & 58 ($\pm$ 6) & 33 ($\pm$ 4)  &  91 ($\pm$ 7)  \\
Q-L3 & 58 ($\pm$ 6) & 33 ($\pm$ 4)  &  91 ($\pm$ 7)  \\
\hline
B-L2 & 70 ($\pm$ 8) & 80 ($\pm$ 12)  &  134 ($\pm$ 9)  \\
B-L3 &  86 ($\pm$ 9) & 104 ($\pm$ 11)  &  111 ($\pm$ 9)  \\
B-L4 &  131 ($\pm$ 11) & 186 ($\pm$ 15)  &  103 ($\pm$ 9)  \\
\hline
1D \cite{rosenhahn2025optimization} & 55 ($\pm$ 6) & 26 ($\pm$ 4)  &  89 ($\pm$ 8)  \\
{\bf ours} & 51 ($\pm$ 6) & 21 ($\pm$ 4)  &  83 ($\pm$ 9)  \\
\hline
\end{tabular}
\caption{Statistics for Quantum circuit optimization  for a nisq architecture (IBM) and comparison of our method to the  qiskit optimizer on levels 1-3 and the BQSKit compiler on optimization levels 2-4 over 100 runs
(8 qubits). The values are the mean and in brackets we provide the standard deviation.}
\label{tab:nisqCompL1}
\end{table}

\begin{table}
\begin{tabular}{|c|c|c|}
\hline
Method & Gate Depth $\downarrow$ &   Time (sec.)$\downarrow$ \\
\hline
Q-L1   & 87($\pm$ 43) &  0.0056\\
Q-L2   & 91($\pm$ 56) & 0.004\\
Q-L3   & 91($\pm$ 56)& 0.008\\
\hline
B-L2   & 87($\pm$ 52)& 5.5\\
B-L3  & 89($\pm$ 75)& 14.3\\
B-L4   &113($\pm$ 187) & 89\\
\hline
& & TTC \\
\hline
1D \cite{rosenhahn2025optimization}&  75($\pm$ 43)& 4200  \\
{\bf ours} & 66($\pm$ 48) &  1400\\
\hline
\end{tabular}
\caption{
Statistics on gate depth and compute time for quantum circuit optimization  for an ion trap architecture and comparison of our method to the  qiskit optimizer on levels 1-3 and the BQSKit compiler on optimization levels 2-4 over 100 runs
(8 qubits). The values are the mean and in brackets we provide the standard deviation. TTC stands for time-to-convergence. It is the last time point where the circuit has been updated. Afterwards no further improvements have been found. 
}
\label{tab:IOCounts}
\end{table}

\begin{table}
\begin{tabular}{|c|c|c|}
\hline
Method & Gate Depth $\downarrow$ &  Time (sec) $\downarrow$ \\
\hline
Q-L1  & 96($\pm$ 41)   & 0.004 \\
Q-L2  & 71($\pm$ 44)  & 0.0037\\
Q-L3  & 71($\pm$ 44) & 0.006\\
\hline
B-L2   &109($\pm$ 76) & 3.2\\
B-L3  &110($\pm$ 72) & 10.6\\
B-L4   &149($\pm$ 284) & 135.9\\
\hline
& & TTC \\
\hline
1D \cite{rosenhahn2025optimization}& 67 ($\pm$ 45)& 3800 \\
{\bf ours} & 63($\pm$ 56) & 1200 \\
\hline
\end{tabular}
\caption{
Statistics on gate depth and compute time for Quantum circuit optimization  for a nisq architecture (IBM) architecture and comparison of our method to the  qiskit optimizer on levels 1-3 and the BQSKit compiler on optimization levels 2-4 over 100 runs
(8 qubits). The values are the mean and in brackets we provide the standard deviation.  As before, TTC stands for time-to-convergence. It is the last time point where the circuit has been updated. Afterwards no further improvements have been found. 
}
\label{tab:nisqCounts}
\end{table}

\begin{figure}[ht]
\centering
\includegraphics[width=0.475\textwidth]{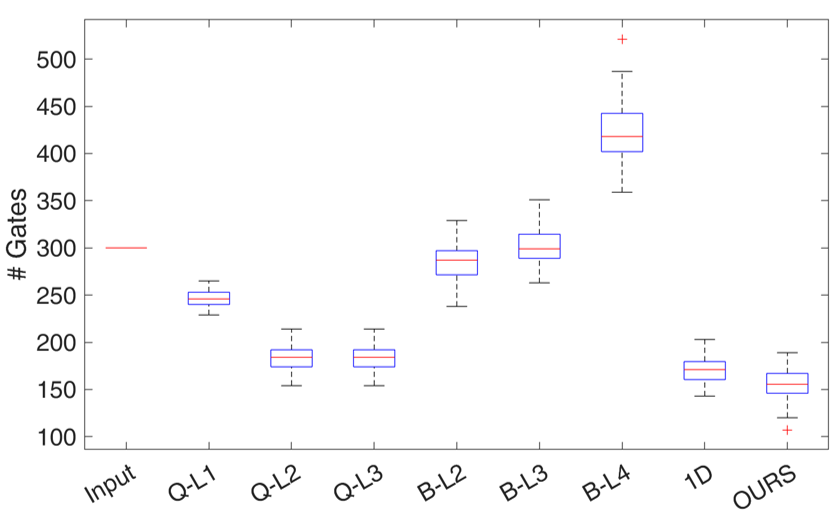}    
\caption{
Statistical summary of the baseline compilation methods qiskit and BQSKit, compared to our proposed method. 
The experiment is similar to Figure \ref{fig:PerfStatIon} using
100 different randomly sampled
quantum circuits of length 300 (8 qubits). 
The gate set is based on a NISQ architecture.
}
\label{fig:PerfStat1}
\end{figure}

Table \ref{tab:IOCompL1} and Figure \ref{fig:PerfStatIon} 
summarize the statistics for quantum circuit optimization  for an ion trap architecture and a comparison of our method to the  qiskit optimizer on levels 1-3 and the BQSKit compiler on optimization levels 2-4 over 100 runs. 
Table \ref{tab:IOCounts} summarizes the statistics on the obtained gate depth and the required compute time to achieve the result  for an ion trap architecture. TTC stands for Time-To-Convergence. It is the timepoint of our algorithm from which on no further improvement has been found until the time budget (1,5 hours) was used up. Our approach is slower, compared to other approaches, but is able to find more efficient realizations in terms of gate count and gate depth. 
The same is shown in
Table \ref{tab:nisqCompL1} and Figure \ref{fig:PerfStat1}
for  a NISQ architecture (IBM). We observed severe performance differences between qiskit and BKSKit across the used gate sets. In contrast, our method is superior and independent of the available gate set. 
It should be noted that BQSKit prioritizes the optimization of the gate count for 2-qubit gates, which explains its subpar performance in terms of total gate count at Level 4, see e.g. Figure~\ref{fig:PerfStatIon}.
Table \ref{tab:nisqCounts} summarizes the statistics on the obtained gate depth and the required compute time to achieve the result for a nisq architecture (IBM) architecture. As before, TTC stands for Time-To-Convergence. It is the timepoint of our algorithm from which on no further improvement has been found until the time budget (1,5 hours) was used up. Again, our approach is slower, compared to other approaches, but is able to find more efficient realizations in terms of gate count and gate depth. The discrepancy of tables \ref{tab:IOComp} and \ref{tab:NIComp} with a short compute time for TTB
with tables \ref{tab:IOCounts} and \ref{tab:nisqCounts} and a long compute time for TTC
can be nicely explained with figures  \ref{fig:ConvNQub} and \ref{fig:ConvNQubNI}: The system has a fast decay in the beginning, where all \textit{easy} replacement rules are exploited. Afterwards, it is getting harder to identify possible algebraic replacements. Please note, that while Figures \ref{fig:ConvNQub} and  \ref{fig:ConvNQubNI}  only show 200 iterations, reaching qiskit level 3 within seconds, the final results were obtained after 30.000 iterations and a much longer compute time.

\subsection{Factorizing numbers}
In 2021 \cite{Vandersypen2001} it had been experimentally shown that quantum computers can factor the number 15. This is possible by a quantum circuit using 6 qubits and around 17 gates, including CNOT, CPHASE and Toffoli gates. Since this implementation was performed on an NMR quantum computer, 
a runner-up to this work has been presented in \cite{doi:10.1126/science.aad9480} for the case of ion trap architectures. 
As discussed in the post \cite{webFac21}, the factorization of the number 21 is much harder and comprises of 15 qubits and around 650 gates, including 191 CNOT gates and 369 Toffoli gates. Based on the provided .qasm-file from \cite{webFac21}, 
a direct transpilation of this circuit to an ion trap architecture (qiskit transpilation, optimization level 0) leads to 16180 gates. It is obvious that decoherence makes it infeasible to run this code on a current real quantum computer, but we selected this toy example to demonstrate that our method can still improve gate counts on a large scale setting. 
\begin{table}
\begin{tabular}{|c|c|c|c|}
\hline
Original & qiskit (lv3) $\downarrow$ &  BQSKit (lv3) $\downarrow$ & {\bf ours} $\downarrow$\\
15 qubits & & & \\
\hline
16179 & 10055 & 7485 & 7129 \\
\hline
\end{tabular}
\caption{Gate count for transpiling the factorization of the number 21 on an ion trap architecture.}
\label{tab:Fac21}
\end{table}

Table \ref{tab:Fac21} summarizes our obtained results and shows the qiskit transpilaton result on optimization level 0, the qiskit optimization result on level 3 and the best result we could obtain with BQSKit (which was on level 3 and after several repetitions). In comparison, our approach based on random sampling could also process and optimize the circuit and we were able to improve on it. However, it should be noted that our implementation had to optimize for several days to achieve this result. 
Thus, in future works we will focus on making our implementation more efficient for upscaling.

\section{Summary and Discussion}
A naive mapping of a quantum circuit to an existing hardware can lead to inefficient quantum circuits with unnecessary redundancies. The increasing decoherence in longer quantum circuits makes it mandatory to optimize for equivalent and shorter circuits.
In this work, we introduced a 2D random search scheme to significantly shorten circuit length while preserving the underlying unitary matrix of the quantum circuit.
The efficient 2D selection of likely reducible sub circuits can be enhanced by using a neural network which predicts a likelihood map of potentially reducible circuit blocks. Thus, neural guided sampling allows for an efficient reduction scheme which leads in general to shorter circuits, compared to earlier presented 1D approaches or standard implementations in qiskit or BQSKit. Additionally, the optimization time can be heavily reduced due to the improved sampling efficiency. Whereas the quality of standard optimizers can be reached within seconds (our measured TTB score), highly efficient quantum circuits are still demanding to optimize, as it is still an NP-hard problem at the end.
In future works, we will also integrate costs for decoherence time of single gates and sub circuits.  E.g. a circuit with six gates and two $CZ$-gates might have a higher decoherence than a nine operator circuit with only one $CZ$ gate. 
Using different cost measures, we plan to update our optimization criteria, which currently focus solely on the number of gates, with alternative metrics in the future. Additionally, we will work on alternative optimization procedures, e.g. based on mixed-integer linear programming with provable optimization guarantees \cite{Bod2023a}.

\subsection*{Competing interest}
The authors have no competing interests.
\subsection*{Author Contributions}
B.R. conceived the initial idea, implemented the methods, and performed the analysis. T.J.O. and C.H. contributed to refining the analysis and developing the applications. All authors contributed to the interpretation of results and to writing the manuscript. B.R. prepared the first draft, and all authors reviewed and approved the final version.
\subsection*{Funding Declaration}
This work was supported, in part, by the Federal Ministry of Research, Technology and Space (BMFTR), Germany under the AI service center KISSKI (grant no. 01IS22093C), the QC service center QUICS (grant no. 13N17418,
by the Quantum Valley Lower Saxony and by Germany's Excellence Strategies EXC-2122 PhoenixD and EXC-2123 Quantum Frontiers and the ERC and the DFG via the project ResourceQ.

\bibliographystyle{plain}
\bibliography{sample}

\end{document}